\documentclass[doublecol]{epl2} 

\title{Identifying effective multiple spreaders by coloring complex networks
}
\shorttitle{Identifying effective multiple spreaders by coloring complex networks} 

\author{Xiang-Yu Zhao\inst{1,2}\and Bin Huang\inst{3}\and Ming Tang\inst{1}\footnote{tangminghuang521@hotmail.com}
\and Hai-Feng Zhang\inst{4,5}\footnote{haifengzhang1978@gmail.com} \and Duan-Bing Chen\inst{1}}
\shortauthor{Xiang-Yu Zhao \etal}
\institute{\inst{1} Web Sciences Center, University of Electronic Science
and Technology of China, Chengdu 611731, P. R. China\\ \inst{2} School of Computer Science and Technology, University of Science and Technology of China, Hefei 230027, P. R. China\\ \inst{3} School of Applied Mathematics, Chengdu University of Information Technology, Chengdu 610225, P. R. China\\ \inst{4} School of Mathematical Science, Anhui University, Hefei 230601, P. R. China\\ \inst{5} Department of Communication Engineering, North University of China, Taiyuan, Shan'xi 030051, P. R. China}

\pacs{89.75.Hc}{Networks and genealogical trees}
\pacs{89.75.Fb}{Structures and organization in complex systems}
\pacs{87.23.Ge}{Dynamics of social systems}

\abstract {How to identify influential nodes in social networks is of theoretical significance, which relates to how to prevent epidemic spreading or cascading failure, how to accelerate information diffusion, and so on. In this Letter, we make an attempt to find \emph{effective multiple spreaders} in complex networks by generalizing the idea of the coloring problem in graph theory to complex networks. In our method, each node in a network is colored by one kind of color and nodes with the same color are sorted into an independent set. Then, for a given centrality index, the nodes with the highest centrality in an independent set are chosen as multiple spreaders. Comparing this approach with the traditional method, in which nodes with the highest centrality from the \emph{entire} network perspective are chosen, we find that our method is more effective in accelerating the spreading process and maximizing the spreading coverage than the traditional method, no matter in network models or in real social networks. Meanwhile, the low computational complexity of the coloring algorithm guarantees the potential applications of our method.}

\begin{document}

\maketitle
\section{Introduction} \label{sec:intro}

Spreading phenomenon is ubiquitous in nature, which describes many
important activities in society~\cite{pastor2014epidemic}. Examples include the propagation of infectious diseases, the dissemination of information (e.g., ideas, rumors, opinions, behaviors), and the diffusion of new technological innovations. With the advancement of complex network theory, spreading dynamics on complex networks have been intensively studied in the past decades. Many studies have revealed that the spreading process is strongly influenced by the network topologies~\cite{Newman:book,PSV:2001b}.

An important issue in analyzing complex networks is to identify the most influential nodes in a spreading process, which is crucial for developing efficient strategies to control epidemic spreading, or accelerate information diffusion. For this reason, more and more attentions have been paid
to identify the  most influential nodes in networks~\cite{kempe2003maximizing,kitsak2010identification,zhang2011node,PhysRevE.85.066123,chen2012identifying,zeng2013ranking,de2014role}. Many centrality indices have been proposed, such as, degree centrality (defined as the degree of a node)~\cite{newman2003structure}, betweenness centrality (measured by the number of times that all shortest paths travel through the node)~\cite{freeman1977set}, eigenvector centrality (defined as the dominant eigenvector of the adjacency matrix)~\cite{estrada2005subgraph}, neighborhood centrality (defined as the average connectivity of all neighbors)~\cite{maslov2002specificity} and closeness centrality (reciprocal of the sum of the lengths of the geodesic distance to every other node)~\cite{sabidussi1966centrality}. Recently Kitsak \emph{et~al}. proposed a \emph{k}-core decomposition to identify the most influential spreaders, which is found to be better than the degree centrality index in many real networks~\cite{kitsak2010identification}. However, most of these methods measure the influence of each node from the viewpoint of entire network, which may be particularly suitable to the case in which single spreader of information is considered (i.e., only one node is selected as the initial spreader)~\cite{SenPei2014,Liu:2014}. Many times, the spreading processes of rumors, ideas, opinions, or advertisements may initiate from different spreaders. In this case, the traditional methods that only select the nodes in the top of one certain ranking (e.g., the ranking obtained by ordering the nodes according to the degree centrality) may be not the optimal strategy since these chosen nodes cannot be dispersively distributed~\cite{hu2014effects}. Thus, how to properly choose the multiple initial spreaders is an important and challenging problem. To this end, apart from the consideration for the influence of each node, we need to make the chosen spreaders being sufficiently dispersive to ensure that the information can quickly diffuse.

Motivated by the above reasons, in this Letter, we propose a method to detect the \emph{effective multiple spreaders} which can enhance the spreading processes. For this method, the independent sets with disjointed nodes are obtained by coloring a network at first, and then the nodes with the highest centrality in an independent set are chosen as the initial spreaders. By implementing extensively simulations on network models as well as the real networks, we find that our method can effectively enhance the spreading process. More importantly, the computational complexity of our method is $O(N^2)$ when the size of network is $N$, which further ensures the potential applicability of this method.

The remainder of this paper is organized as follows. In Sec.~\uppercase\expandafter{\romannumeral2}, we first describe the details of our method. Then we present the main results in Sec.~\uppercase\expandafter{\romannumeral3}. Finally, we summarize the conclusions in Sec.~\uppercase\expandafter{\romannumeral4}.

\section{Methods} \label{sec:model}

The four-color theorem is the most famous theorem in the graph coloring problem, which states that, given any a plane graph, no more than four colors are required to color the regions of the plane graph so that no two adjacent regions have the same color~\cite{bollobas1998modern,appel1977every1,appel1977every2}. In other words, the number of colors for all vertexes in a plane graph is not greater than four. The graph coloring problem has a wide range of applications, including the problem of the wireless channel allocation~\cite{riihijarvi2005frequency}, the problem of the scheduling~\cite{leighton1979graph,welsh1967upper}, and so on. Here, we generalize the idea of the graph vertex coloring problem to complex networks to obtain the \emph{effective multiple spreaders}.
The main steps are as follows. We first color a given network $G=(V, E)$ ($V$ denotes the set of nodes and $E$ denotes the set of edges) with the Welsh-Powell algorithm~\cite{welsh1967upper}~[see below] and each node in set $V$ is colored by one kind of color. Secondly, sorting the nodes with the same color into the same subset $V_i, i=1,2,\cdots,K$ (each subset is called an independent set, $K$ denotes the number of colors being used to color the network), which ensures that $V=V_1\cup V_2\cdots\cup V_K$ and $V_i\cap V_j=\phi, \forall i\neq j$, where $\phi$ is an empty set. As nodes with the same color are not directly connected, the distance between any two nodes in an independent set will not be smaller than two. Lastly, we choose the nodes with highest centrality index in an independent set as the initial spreaders. To ensure that there are sufficient nodes from which to choose, we give priority to those large independent sets with more nodes, especially \emph{the largest independent set} with the maximum nodes in this Letter.

Though there are many graph coloring algorithms, an ideal algorithm should have the qualities: the time complexity is low and the number of colors to color the network is few since many real networks have huge sizes. In view of this, we choose the Welsh-Powell algorithm, whose time complexity is $O(N^2)$~\cite{klotz2002graph}. For a given  network $G=(V, E)$ with $V=\{v_1,v_2,\cdots,v_N\}$, we let the color function be $\pi$ and the color set be $C=\{1, 2,\cdots,\Delta+1\}$,  where $\Delta$ be the maximum degree of network $G$. The details of the Welsh-Powell algorithm are~\cite{welsh1967upper}: \\

\textbf{Step 1:} according to the degree centrality, re-rank the node set $V$ in descending order, such that $k(v_1)\geq k(v_2)\geq\cdots\geq k(v_N)$;\\

\textbf{Step 2:} let $\pi(v_1)=1$, $i=1$;\\

\textbf{Step 3:} if $i=N$, stop; otherwise, let $C(v_{i+1})=\{\pi(v_j)|j\leq i$,~and~$v_j$~is~connected~by~ $v_{i+1}\}$. Let $m$ be the minimal positive integral of the subset $C\setminus C(v_{i+1})$ [where $C\setminus C(v_{i+1})$ is the complementary set of the subset $C(v_{i+1})$ in the set $C$]. Then $\pi(v_{i+1})=m$.\\

\textbf{Step 4:} let $i=i+1$, and back to step 3.\\

In the above algorithm, $k(v_i)$ denotes the degree of node $v_i$, and $\pi(v_i)=m$ denotes that the node $v_i$ is colored by a color labeled $m$.

When the multiple spreaders are selected, a spreading model should be used to check the effectiveness of the proposed method. In many literatures, the susceptible--infected--recovered (SIR) epidemic model is used to simulate the spreading process in networks, in which each node can be in one of three states: susceptible, infected, or recovered. A susceptible node is healthy and can catch the disease from each infected neighbor with transmission rate $\beta$, whereas an infected node becomes recovered with recovery rate $\mu$ and is immune to the disease. In the classical SIR model, each infected node can contact \emph{all} of its neighbors at per time step with transmission rate $\beta$. In reality, at a time step, one often can contact \emph{one} neighbor at most, taking the sex activity and the telephone marketing activity as examples. Thus, in this Letter, we use the SIR epidemic model based on a contact process to simulate the spreading process and measure the effectiveness of the proposed method~\cite{Castellano:2010,yang2008selectivity,PhysRevE.85.026116}. It’s worth noting that our method can also be applied to the classical SIR model. The epidemic spreading process ends when there is no infected node in the network. We define the effective transmission rate $\lambda=\beta/\mu$ by fixing the recovery rate $\mu=0.1$.

\section{Results} \label{sec:main results}
One should note that our method is compared with the traditional method according to one given centrality index. Taking the degree centrality index as an example, for our method, the nodes in \emph{the largest independent set} are ranked in a descending order according to the degree centrality index, and then the nodes at the top of the ranking are selected as spreaders (labeled as IS method). For the traditional method, all nodes are ranked according to the degree centrality index from \emph{entire network} perspective, and the same amount nodes at the top of the ranking are selected as the spreaders (labeled as EN method).

To measure the effectiveness of the IS method, we define the relative difference of outbreak size $\Delta r_R$ as $\Delta r_R=(R_i-R_e)/R_e$, where $R_i$ and $R_e$ are the final number of recovered nodes for the IS method and the EN method, respectively. Thus, the larger value of $\Delta r_R$ is, the better effectiveness of the IS method is. All results are averaged over $500$ independent realizations.

We first perform the Welsh-Powell algorithm to Barab\'{a}si-Albert (BA) network with size $N=10000$ and average degree $\langle k\rangle=12$~\cite{barabasi1999emergence}. From the inset of Fig.~1(a), one can see that such a network can be successfully divided into different independent sets, where the number of color is $K=8$ and the node number in the largest independent set is over $2000$. Fig.~1 also compares the IS method with the EN method based on the degree centrality index. In general, $\Delta r_R$ is larger than $0$ for the different initial spreaders $n_0$ and the different values of transmission rate $\lambda$. This indicates that the IS method is better than the latter case in the most situations. More importantly, from Fig.~1(a) one can see that the advantage of the IS method is the most striking when $\lambda$ is not too small or too large. As we know, when $\lambda$ is very small, the information initiated from any node can only spread to a very small fraction of nodes. The influence regions of multiple spreaders for these two methods scarcely overlap each other. The outbreak size is approximately equal to the sum of multiple spreaders' spreading coverage (i.e., the number of infected nodes). Thus, the difference of the two methods cannot be distinguished obviously at a small $\lambda$, which results in a small value of $\Delta r_R$. With the increase of transmission rate $\lambda$, single node can induce a greater spreading coverage. More dispersive locations of multiple spreaders for the IS method lead to less overlap of the influence regions, and the IS method thus performs better. When $\lambda$ is very large, the information can diffuse to a very wide range even single node is selected as the initial spreader. In this case, there are too many overlap influence regions to play the role of the IS method. In Fig.~1(b), $\Delta r_R$ as a function of $n_0$ displays distinct trends for different values of $\lambda$, which stems from the combined effects of both the dispersive locations of multiple spreaders and the intricate spreading processes at different $\lambda$. These distinct trends will be verified by the relative difference of effective contacts in Fig.~2(c) later.

\begin{figure}
\begin{center}
\includegraphics[width=3in]{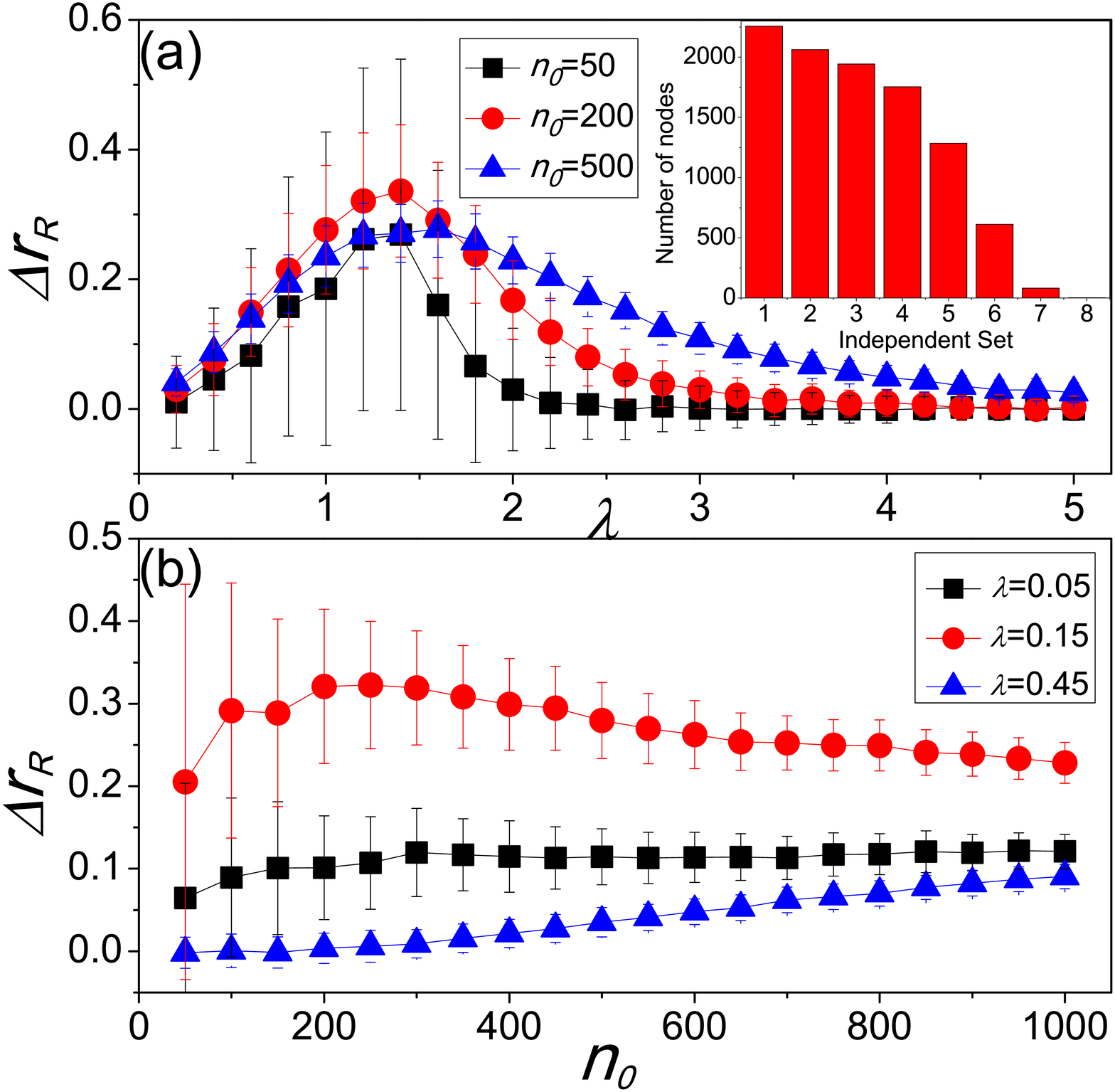}
\caption{(Color online) For the degree centrality index, the EN method and the IS method are compared in the BA network. (a) the relative difference of outbreak size $\Delta r_R$ as a function of the effective transmission rate $\lambda$ for different number of initial spreaders $n_0$; (b) $\Delta r_R$ as a function of $n_0$ for different values of $\lambda$. The inset of subfigure (a) is the size distribution of independent sets.The error bars are given by the standard deviation.}
\label{fig5}
\end{center}
\end{figure}

To maximize the spreading coverage, we hope these spreaders not only have high centrality, but also are dispersive enough so that a susceptible node has only one or few infected neighbor instead of surrounded by many infected nodes to reduce the overlap effect of the spreaders. As a result, if we can verify that the initial spreaders in the IS method have a larger average distance among them and produce less overlap of spreading influence (i.e., more effective contacts between infected nodes and their susceptible neighbors) than that in the EN method, the phenomena in Fig.~1 will be naturally explained.
For this reason, we define two metrics. One is the relative difference of average distance, $\Delta r_D=(D_i-D_e)/D_e$, where $D_i$ and $D_e$ are the average distance among the initial spreaders based on the IS method and the EN method, respectively. The other is the relative difference of effective contacts, $\Delta r_C=(C_{i}-C_{e})/C_{e}$, where $C_{i}$ and $C_{e}$ are the number of effective contacts between infected nodes and their susceptible neighbors based on the IS method and the EN method, respectively. In each time step of the transmission processes, an infected node randomly chooses one of its neighbors to transmit the information with probability $\beta$. If the chosen neighbor is susceptible, this contact is defined as an effective contact; otherwise, it is not effective. A greater number of effective contacts denotes the less overlap of spreading regions initiated from multiple spreaders. In Fig.~2, one can see that $\Delta r_D$ is always larger than $0$ [see Fig.~2(a)]. The reason can be explained as follow: the distance between any two nodes in an independent set is greater than or equal to $2$, while the nodes with highest centrality are connected more easily in the BA network. In Figs.~2(b) and~2(c), $\Delta r_C$ is generally larger than $0$, too. Moreover, comparing Figs.~2(b) and~(2)(c) with Figs.~1(a) and~1(b), respectively, one can observe that the effects of $\lambda$ and $n_0$ on $\Delta r_D$ and $\Delta r_C$ are similar to that on $\Delta r_R$. Thus, the reason for the advantage of the IS method is well explained. To be specific, the greater $\Delta r_D$ induces the greater $\Delta r_C$ and then results in the greater $\Delta r_R$.
\begin{figure}
\begin{center}
\includegraphics[width=3in]{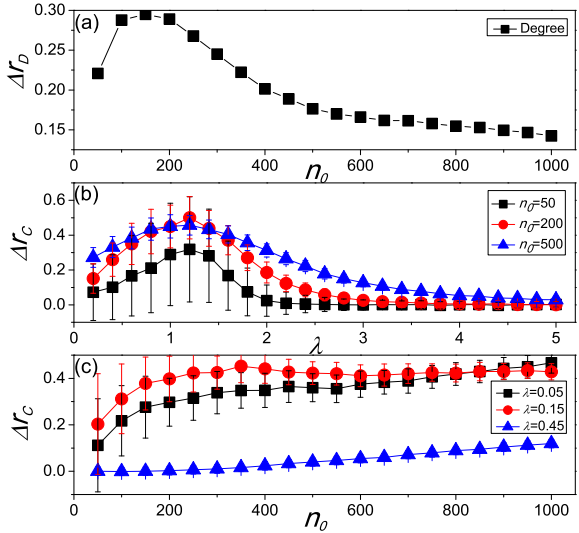}
\caption{(Color online) For the degree centrality index, the relative difference of average distance $\Delta r_D$ and the relative difference of effective contacts $\Delta r_C$ are plotted to explain the phenomena in Fig.~1. (a) $\Delta r_D$ vs. $n_0$, (b) $\Delta r_C$ vs. $\lambda$ for different values of $n_0$, (c) $\Delta r_C$ vs. $n_0$ for different values of $\lambda$. The detail definitions of the two metrics are given in the main text.}
\label{fig6}
\end{center}
\end{figure}

\emph{}In Fig.~3, we further compare the IS method with the EN method based on five commonly used indices--degree centrality, betweeness centrality, closeness centrality, eigenvector centrality, and neighborhood centrality. Since the \emph{k}-core decomposition can not quantify the relative influence of nodes in the BA network, this index is not considered here~\cite{zeng2013ranking}. As in Fig.~1, the results in Fig.~3 indicate that the IS method is more effective than the EN method for all cases. In particular, the advantage of the IS method is the most remarkable for the betweeness centrality index. The positive $\Delta r_D$ and $\Delta r_C$ for different cases shown in Fig.~4 can explain that the IS method is superior to the EN method for different centrality indices. Meanwhile, Figs.~4(a) and~4(b) show that the methods based on the degree and betweeness centrality indices can generate the largest values of $\Delta r_D$ and $\Delta r_C$, leading to the highest efficiency in enhancing the spreading coverage in Fig.~3.

\begin{figure}
\begin{center}
\includegraphics[width=3in]{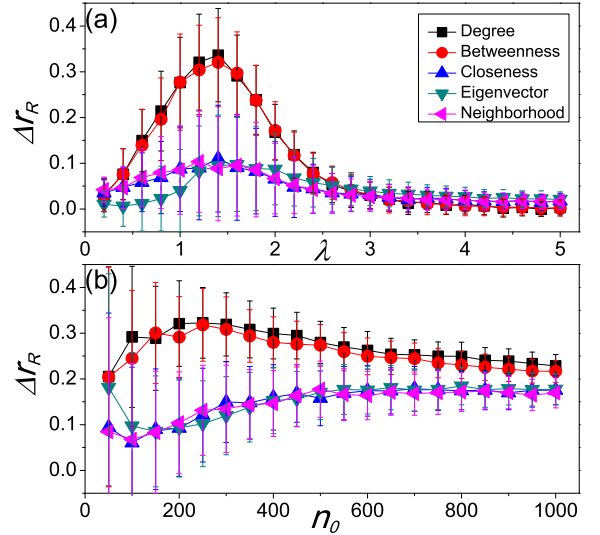}
\caption{(Color online) Comparing the IS method with the EN method based on five indices (i.e., degree, betweeness, closeness, eigenvector, and neighborhood) in the BA network. (a) $\Delta r_R$ vs. $\lambda$ at $n_0=200$, (b) $\Delta r_R$ vs. $n_0$ at $\lambda=1.5$.}
\label{fig5}
\end{center}
\end{figure}

\begin{figure}
\begin{center}
\includegraphics[width=3in]{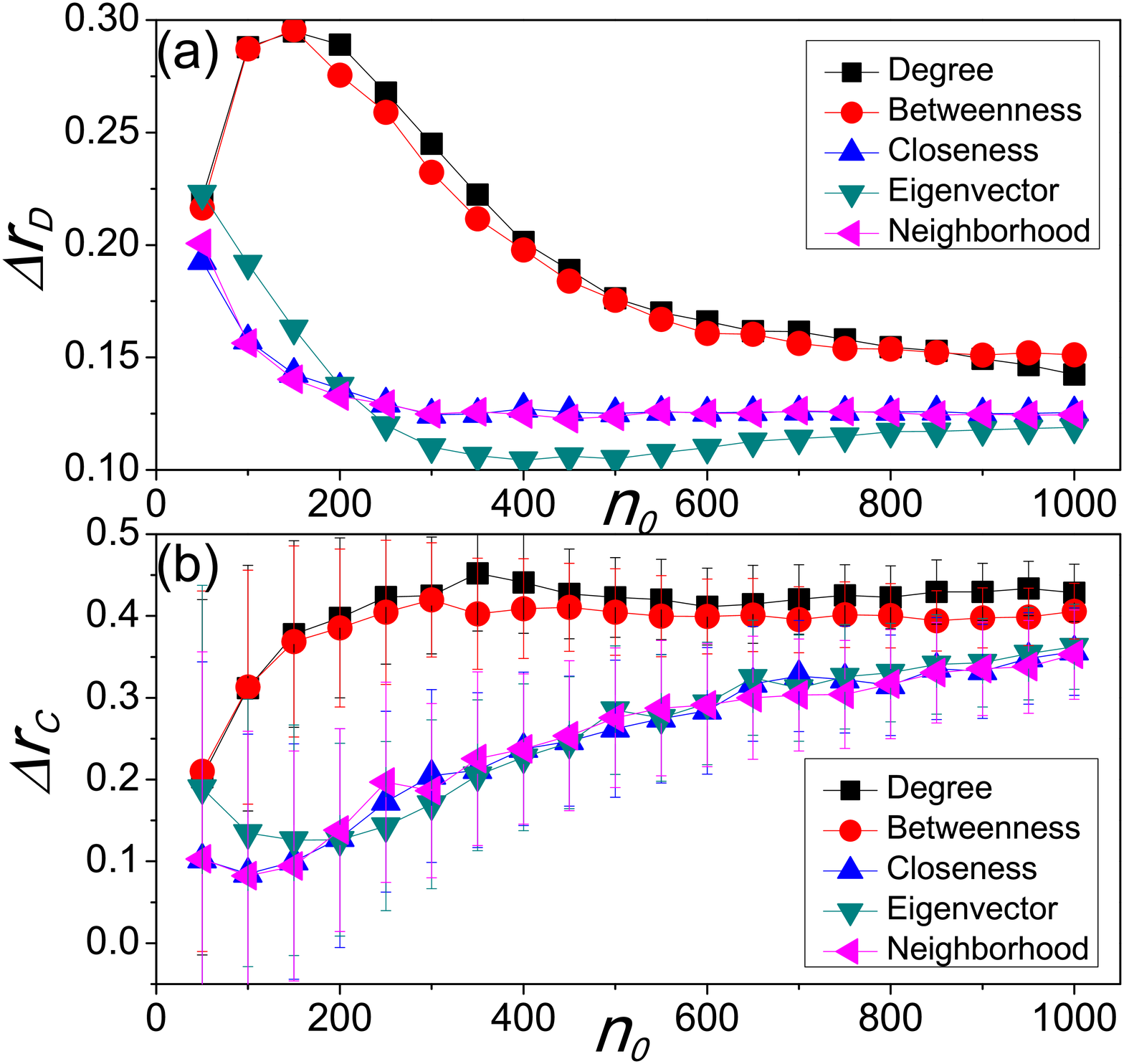}
\caption{(Color online) $\Delta r_D$ and $\Delta r_C$ as the functions of $n_0$ are given to explain the phenomena in Fig.~3. (a) $\Delta r_D$ vs. $n_0$, (b) $\Delta r_C$ vs. $n_0$. The parameter is chosen as $\lambda=1.5$.}
\label{fig5}
\end{center}
\end{figure}

The time evolutions of the relative difference of outbreak size $\Delta r_R(t)=[R_i(t)-R_e(t)]/R_e(t)$ for different indices are also shown in Fig.~5. As shown in Figs.~5(a) [$n_0=200$ and $\lambda=0.15$] and~5(b) [$n_0=500$ and $\lambda=0.3$], the value of $\Delta r_R(t)$ is generally larger than 0. It means that, compared with the EN method, the IS method can not only extend the spreading coverage but also \emph{speed up} the spreading process. Moreover, Fig.~5(a) illustrates that $\Delta r_R(t)$ monotonously increases with time step $t$ when the values of $n_0$ and $\lambda$ are small. Nonetheless, when the values of $n_0$ and $\lambda$ are large, $\Delta r_R(t)$ increases with $t$ at first and then decreases to a stable level [see Fig.~5(b)]. For the latter case, the information begins to diffuse from the initial spreaders in the early stages, the IS method ensures these multiple spreaders are more dispersive, leading to the better effectiveness of the IS method. With the further increase of $t$, the information will diffuse to a wide range of network and the influence regions of the multiple spreaders are more likely to overlap each other, the advantage of the IS method will thus be weakened.
\begin{figure}
\begin{center}
\includegraphics[width=3in]{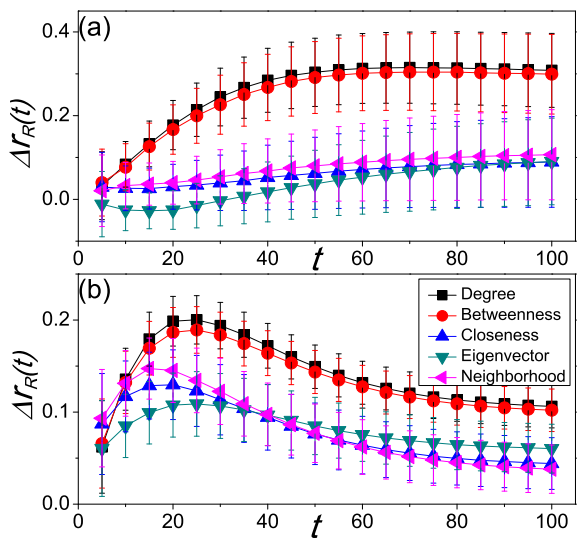}
\caption{(Color online) For five indices, the time evolutions of $\Delta r_R(t)$ in the BA network are plotted for (a) $\lambda=1.5$ and $n_0=200$, (b) $\lambda=3.0$ and $n_0=500$.}
\label{fig5}
\end{center}
\end{figure}

Finally, we use two real networks--Blogs network~\cite{xie2006social} and Email network~\cite{guimera2003self}. To further confirm the effectiveness of our method, where some basic structural features of the two networks are given in Table~\ref{tab:1}. In Fig.~6, $\Delta r_R$ as a function of $\lambda$ and $n_0$ for six centrality indices are presented. Note that \emph{k}-core index is also considered, besides the five indices shown in Fig.~3. No matter the Blogs network [Figs.~6(a) and 6(b)] or the Email network [Figs.~6(c) and 6(d)], the IS method is more effective in enhancing the spreading process than the EN method. In order to further verify our method, other independent sets such as the second largest one are also investigated. As expected, all simulations reveal the same conclusion.

\begin{table}
\centering
 \caption{\label{tab:1} Basic structural parameters of Blogs network and Email network. $N$ is the total number of nodes, $\langle k\rangle$ denotes the
average degree, and $H$ is the degree heterogeneity, defined as $H =\langle k^2\rangle/\langle k\rangle^2$. $D$ is the average shortest distance, and $C$ and $r$ are the clustering coefficient and assortative coefficient, respectively.}

\begin{tabular}{llllllll}
\hline
 Network&  $N$ & $\langle k\rangle$ & $H$ & $D$& $C$ & $r$    \\ \hline  
  Blogs & 3982 & 3.42& 4.038&6.227&	0.146&	0.133 \\         
   Email & 1133& 9.62	&1.942 &3.716	&0.110& 0.078\\ \hline
\end{tabular}
\end{table}

\begin{figure}
\begin{center}
\includegraphics[width=3.3in]{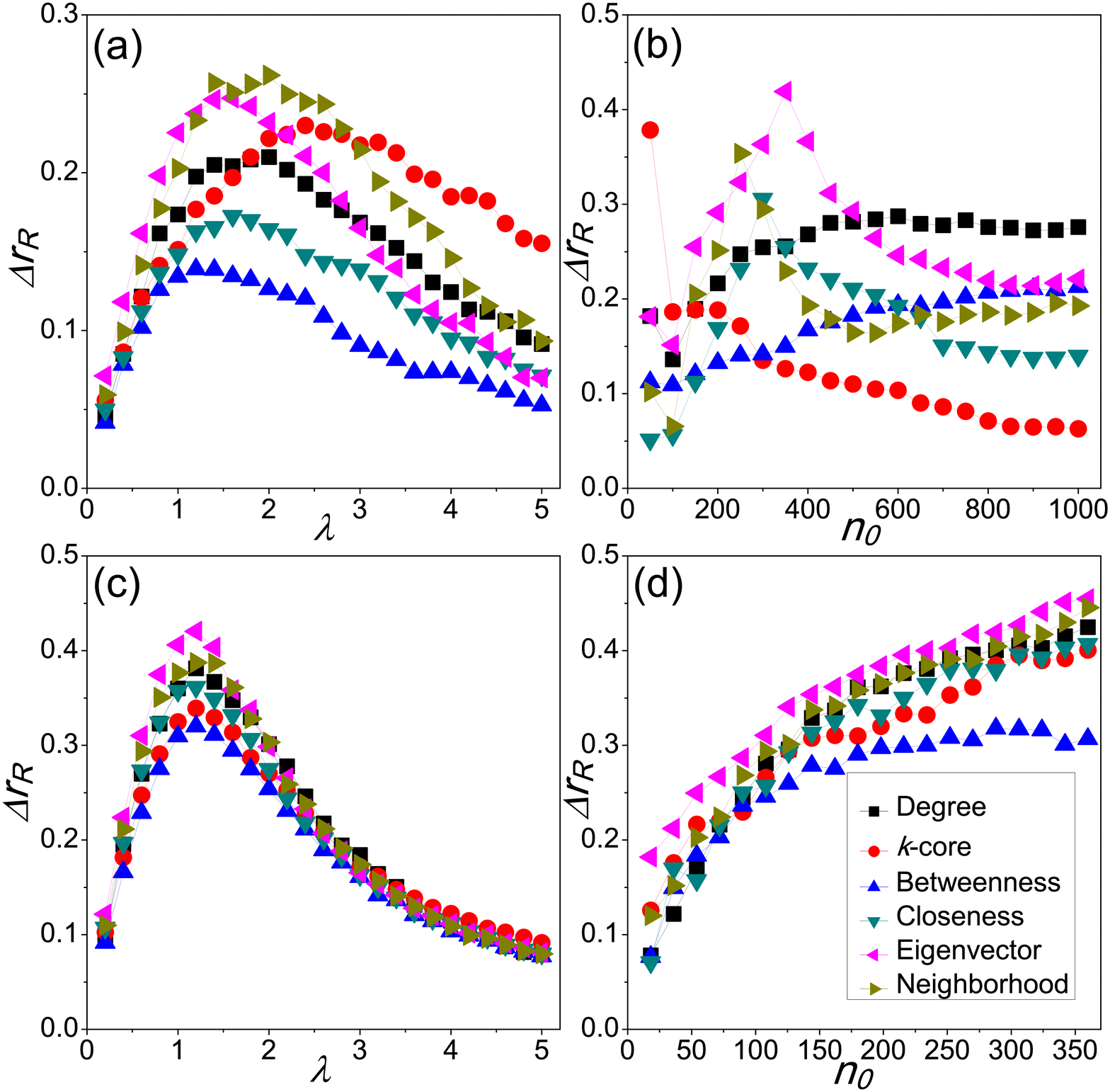}
\caption{(Color online) For different indices, $\Delta r_R$ in Blogs and Email networks are given as the functions of $\lambda$ and $n_0$, respectively. $\Delta r_R$ as a function of the values $\lambda$ (a) and $n_0$ (b), respectively in Blogs network; $\Delta r_R$ as a function of the values $\lambda$ (c) and $n_0$ (d), respectively in Email network. Detail network information is summarized in Table~\ref{tab:1}. }
\label{fig5}
\end{center}
\end{figure}

\section{Summary} \label{sec:discussion}

Even though great improvement has been made in the research of identifying influential spreaders, there are still many problems needed to be solved, among which how to find multiple effective spreaders is an important question. It is commonly believe that, to effectively speed up the spreading process, the selected multiple spreaders should be as dispersive as possible to reduce the overlap of the spreading regions initiated from multiple spreaders. But, how to design an effective method to achieve this goal is almost vacant. In this Letter, we have proposed an effective method by selecting the nodes with the highest centrality from an independent set as the initial spreaders rather than from the entire network. By testing such a method on BA network and two real networks, we found that our method can greatly enhance the average distance among these initial spreaders and the effective contacts between the susceptible nodes and the infected nodes. Therefore, our method can ensure that the information diffuses much wider and faster. Meanwhile, the computational complexity of the coloring algorithm used in the paper is $O(N^2)$, which further guarantees its possible applications. Although the efficiency of our method was studied from the perspective of the spreading process, it is immediately related to many other aspects~\cite{huang:218902}, including network resilience to attacks, immunization of epidemics, commercial product promotions in markets and other aspects, which implies the potential applications of our method.

\begin{acknowledgments}
X.-Y. Zhao would like to thank Kai Qi for stimulating discussions. This work was partially supported by the National Natural Science Foundation of China (Grant Nos.~11105025, 61473001, 11331009, 61433014) and China Postdoctoral Science Special Foundation (Grant No. 2012T50711).
\end{acknowledgments}
\bibliographystyle{eplbib}
\bibliography{color}

\begin{thebibliography}{10}
\expandafter\ifx\csname url\endcsname\relax\def\url#1{\texttt{#1}}\fi

\bibitem{pastor2014epidemic}
\Name{Pastor-Satorras R., Castellano C., Van~Mieghem P. \and Vespignani A.}
  \REVIEW{arXiv:1408.2701}{}{2014}{}.

\bibitem{Newman:book}
\Name{Newman M. E.~J.} \Book{Networks - An Introduction} (Oxford University
  Press, New York) 2010.

\bibitem{PSV:2001b}
\Name{Pastor-Satorras R. \and Vespignani A.} \REVIEW{Phys. Rev.
  E}{63}{2001}{066117}.

\bibitem{kempe2003maximizing}
\Name{Kempe D., Kleinberg J. \and Tardos {\'E}.} \Book{Maximizing the spread of
  influence through a social network} in proc. of \Book{Proceedings of the 9th
  ACM SIGKDD international conference on Knowledge discovery and data mining}
  (ACM) 2003 pp. 137--146.

\bibitem{kitsak2010identification}
\Name{Kitsak M., Gallos L.~K., Havlin S., Liljeros F., Muchnik L., Stanley
  H.~E. \and Makse H.~A.} \REVIEW{Nat. Phys.}{6}{2010}{888}.

\bibitem{zhang2011node}
\Name{Zhang J., Xu X.-K., Li P., Zhang K. \and Small M.}
  \REVIEW{CHAOS}{21}{2011}{016107}.

\bibitem{PhysRevE.85.066123}
\Name{Borge-Holthoefer J., Rivero A. \and Moreno Y.} \REVIEW{Phys. Rev.
  E}{85}{2012}{066123}.

\bibitem{chen2012identifying}
\Name{Chen D., L{\"u} L., Shang M.-S., Zhang Y.-C. \and Zhou T.}
  \REVIEW{Physica A}{391}{2012}{1777}.

\bibitem{zeng2013ranking}
\Name{Zeng A. \and Zhang C.-J.} \REVIEW{Phys. Lett. A}{377}{2013}{1031}.

\bibitem{de2014role}
\Name{de~Arruda G.~F., Barbieri A.~L., Rodr{\'\i}guez P.~M., Rodrigues F.~A.,
  Moreno Y. \and Costa L. d.~F.} \REVIEW{Phys. Rev. E}{90}{2014}{032812}.

\bibitem{newman2003structure}
\Name{Newman M. E.~J.} \REVIEW{SIAM Rev.}{45}{2003}{167}.

\bibitem{freeman1977set}
\Name{Freeman L.~C.} \REVIEW{Sociometry}{40}{1977}{35}.

\bibitem{estrada2005subgraph}
\Name{Estrada E. \and Rodriguez-Velazquez J.~A.} \REVIEW{Phys. Rev.
  E}{71}{2005}{056103}.

\bibitem{maslov2002specificity}
\Name{Maslov S. \and Sneppen K.} \REVIEW{Science}{296}{2002}{910}.

\bibitem{sabidussi1966centrality}
\Name{Sabidussi G.} \REVIEW{Psychometrika}{31}{1966}{581}.

\bibitem{SenPei2014}
\Name{Pei S., Muchnik L., Jos{\'e} S.~Andrade J., Zheng Z. \and Makse H.~A.}
  \REVIEW{Sci. Rep.}{4}{2014}{5547}.

\bibitem{Liu:2014}
\Name{Liu Y., Tang M., Zhou T. \and Do Y.} \REVIEW{arXiv:1409.5187}{}{2014}{}.

\bibitem{hu2014effects}
\Name{Hu Z.-L., Liu J.-G., Yang G.-Y. \and Ren Z.-M.} \REVIEW{Europhys.
  Lett.}{106}{2014}{18002}.

\bibitem{bollobas1998modern}
\Name{Bollob{\'a}s B.} \Book{Modern graph theory} Vol. 184 (Springer) 1998.

\bibitem{appel1977every1}
\Name{Appel K., Haken W. \etal} \REVIEW{Illinois J. Math.}{21}{1977}{429}.

\bibitem{appel1977every2}
\Name{Appel K., Haken W., Koch J. \etal} \REVIEW{Illinois J.
  Math.}{21}{1977}{491}.

\bibitem{riihijarvi2005frequency}
\Name{Riihij{\"a}rvi J., Petrova M. \and M{\"a}h{\"o}nen P.} \Book{Frequency
  allocation for {WLAN}s using graph colouring techniques.} in proc. of
  \Book{WONS} Vol.~5 2005 pp. 216--222.

\bibitem{leighton1979graph}
\Name{Leighton F.~T.} \REVIEW{Nat. Bur. Standard}{84}{1979}{489}.

\bibitem{welsh1967upper}
\Name{Welsh D.~J. \and Powell M.~B.} \REVIEW{Comput. J.}{10}{1967}{85}.

\bibitem{klotz2002graph}
\Name{Klotz W.} \REVIEW{Mathematics Report}{5}{2002}{1}.

\bibitem{Castellano:2010}
\Name{Castellano C. \and Pastor-Satorras R.} \REVIEW{Phys. Rev.
  Lett.}{96}{2010}{038701}.

\bibitem{yang2008selectivity}
\Name{Yang R., Huang L. \and Lai Y.-C.} \REVIEW{Phys. Rev.
  E}{78}{2008}{026111}.

\bibitem{PhysRevE.85.026116}
\Name{Borge-Holthoefer J. \and Moreno Y.} \REVIEW{Phys. Rev.
  E}{85}{2012}{026116}.

\bibitem{barabasi1999emergence}
\Name{Barab{\'a}si A.-L. \and Albert R.} \REVIEW{Science}{286}{1999}{509}.

\bibitem{xie2006social}
\Name{Xie N.} \Book{Social network analysis of blogs} Ph.D. thesis MSc
  Dissertation. University of Bristol (2006).

\bibitem{guimera2003self}
\Name{Guimera R., Danon L., Diaz-Guilera A., Giralt F. \and Arenas A.}
  \REVIEW{Phys. Rev. E}{68}{2003}{065103}.

\bibitem{huang:218902}
\Name{Huang B., Zhao X.-Y., Qi K., Tang M. \and Do Y.} \REVIEW{Acta Phys.
  Sin.}{62}{2013}{218902}.

\end{thebibliography}
\end{document}